\documentclass[twocolumn,twoside]{revtex4}

\usepackage{graphicx}
\usepackage{fancyhdr}
\usepackage{amsmath}
\usepackage{xspace}

\newcommand{\CP}{\ensuremath{C\!P}\xspace}
\newcommand{\dztokk}{\ensuremath{D^0 \to K^+ K^-}\xspace}
\newcommand{\dztopipi}{\ensuremath{D^0 \to \pi^+ \pi^-}\xspace}
\newcommand{\invfb}{\xspace fb$^{-1}$\xspace}
\newcommand{\ACPf}{\ensuremath{A_{\CP}(f)}\xspace}
\newcommand{\Adir}{\ensuremath{a^{\mathrm{dir}}_{\CP}(f)}\xspace}
\newcommand{\AdirKK}{\ensuremath{a^{\mathrm{dir}}_{\CP}(K^+K^-)}\xspace}
\newcommand{\Adirpipi}{\ensuremath{a^{\mathrm{dir}}_{\CP}(\pi^+\pi^-)}\xspace}
\newcommand{\Agamma}{\ensuremath{A_\Gamma(f)}\xspace}
\newcommand{\Arawfprompt}{\ensuremath{A^{\pi\mathrm{-tagged}}_{\mathrm{raw}}(f)}\xspace}
\newcommand{\ArawfSL}{\ensuremath{A^{\mu\mathrm{-tagged}}_{\mathrm{raw}}(f)}\xspace}
\newcommand{\ArawKK}{\ensuremath{A_{\mathrm{raw}}(K^+K^-)}\xspace}
\newcommand{\Arawpipi}{\ensuremath{A_{\mathrm{raw}}(\pi^+\pi^-)}\xspace}
\newcommand{\DACP}{\ensuremath{\Delta A_{\CP}\xspace}}
\newcommand{\DACPprompt}{\ensuremath{\Delta A^{\pi\mathrm{-tagged}}_{\CP}\xspace}}
\newcommand{\DACPSL}{\ensuremath{\Delta A^{\mu\mathrm{-tagged}}_{\CP}\xspace}}
\newcommand{\ArawDpKspi}{\ensuremath{A_\mathrm{raw}^{D^+ \to K^0_s\pi^+}}}
\newcommand{\ArawDspKspi}{\ensuremath{A_\mathrm{raw}^{D^{+}_{s} \to K^0_s\pi^+}}}
\newcommand{\ArawDpKsK}{\ensuremath{A_\mathrm{raw}^{D^+ \to K^0_s K^+}}}
\newcommand{\ArawDspKsK}{\ensuremath{A_\mathrm{raw}^{D^{+}_{s} \to K^0_s K^+}}}
\newcommand{\ArawDpphipi}{\ensuremath{A_\mathrm{raw}^{D^+ \to \phi \pi^+}}}
\newcommand{\ArawDspphipi}{\ensuremath{A_\mathrm{raw}^{D^{+}_{s} \to \phi \pi^+}}}
\newcommand{\ACPDspKspi}{\ensuremath{A_{\CP}^{D^{+}_{s} \to K^0_s\pi^+}}}
\newcommand{\ACPDpKsK}{\ensuremath{A_{\CP}^{D^{+} \to K^0_s K^+}}}
\newcommand{\ACPDpphipi}{\ensuremath{A_{\CP}^{D^{+} \to \phi \pi^+}}}

\begin{document}

\title{Charm mixing and CPV}

%

\author{F. Ferrari\\
on behalf of the LHCb collaboration, \\
with results from the Belle collaboration}
\affiliation{University of Bologna and INFN, Bologna, Italy}

\begin{abstract}
In these proceedings, recent results on time-dependent and time-integrated measurements of \CP violation and of meson mixing in the charm sector are presented, including the first observation of \CP violation in the charm system.
\end{abstract}

\maketitle

\thispagestyle{fancy}



\section{Introduction}
Charm mesons provide a unique opportunity to search for \CP violation in decays of particle containing up-type heavy quarks. Due to the large number $D$ mesons collected by the LHCb experiment during the period from 2011 to 2018, an unprecedented experimental precision can be reached.
In these proceedings the latest results published by the LHCb and Belle collaborations concerning the search for \CP violation in charm mesons decays are presented.

\section{Observation of \boldmath{\CP} violation in charm decays} \label{sec:DACP}
Neglecting terms of size $\mathcal{O}(10^{-4})$ and smaller, the \CP asymmetry of a final state $f$, where $f=K^+K^-$ or $\pi^+ \pi^-$, can be written as~\cite{Aaltonen:2011se,Gersabeck:2011xj}
\begin{equation} \label{eq:ACP}
\ACPf \approx \Adir - \frac{\langle t(f) \rangle}{\tau(D^0)} \Agamma
\end{equation}
where $\langle t(f) \rangle$ indicates the mean decay time of $D^0 \to f$ decays in the reconstructed sample, \Adir is the direct \CP asymmetry, $\tau(D^0)$ is the $D^0$ lifetime and \Agamma is the asymmetry between $D^0 \to f$ and $\overline{D}^0 \to f$ effective decay widths~\cite{Aaij:2015yda,Aaij:2017idz}.

The $D^0$ mesons considered in this analysis are produced either in $D^{*+} \to D^0 \pi^+$ decays where the $D^{*+}$ is produced at the primary vertex of the event, referred to as prompt, or in $\overline{B} \to D^0 \mu^- \overline{\nu}_\mu X$ semileptonic decays (charge-conjugation is implied throughout these proceedings), where $\overline{B}$ stands for a hadron containing a $b$ quark and $X$ indicates any additional particle. The flavor of the $D^0$ is inferred from the charge of the accompanying pion $(\pi\mathrm{-tagged})$ in the prompt case and from that of the muon $(\mu\mathrm{-tagged})$ in the semileptonic decay.

The raw asymmetries measured for both types of $D^0$ production mechanisms, \Arawfprompt and \ArawfSL, are defined as the difference between the signal yields of decays tagged by positively or negatively charged pions or muons normalized to their sum. This quantity can be written as
\begin{eqnarray}
\Arawfprompt &\approx& \ACPf + A_\mathrm{D}(\pi) + A_\mathrm{P}(D^*), \nonumber \\
\ArawfSL &\approx& \ACPf + A_\mathrm{D}(\mu) + A_\mathrm{P}(B),
\end{eqnarray}
where $A_\mathrm{D}$ and $A_\mathrm{P}$ represent the detection and production asymmetries of the given particles, respectively. These asymmetries are independent of the final state $f$, and thus cancel in the difference (provided that the kinematic distributions of the relevant particles are equal between the two decay modes), giving
\begin{eqnarray} \label{eq:DACP}
\DACP &\equiv& A_{\CP}(K^+K^-) - A_{\CP}(\pi^+\pi^-) \nonumber \\
&=& \ArawKK - \Arawpipi \nonumber \nonumber \\
&\approx& \Delta \Adir - \frac{\Delta \langle t(f) \rangle}{\tau(D^0)} A_\Gamma,
\end{eqnarray}
where $A_\Gamma$ has been assumed independent of the final state~\cite{Grossman:2006jg,Kagan:2009gb,Du:2006jc}, $\Delta \Adir \equiv \AdirKK - \Adirpipi$ and $\Delta \langle t(f) \rangle$ is the difference of the mean decay times.

The dataset used corresponds to an integrated luminosity of 5.9\invfb collected by the LHCb experiment during 2015-2018. The data are selected in several steps. Requirements are applied on the hardware trigger decision. Fiducial requirements are imposed in order to exclude kinematic regions which have very large raw asymmetries (up to 100\%) due to large detection asymmetries. In the prompt sample a requirement is imposed to suppress the background coming from non-prompt $D^0$ mesons. Particle identification (PID) cuts are also applied to suppress the background due to the mis-identification of final-state particles. The $D^{*+}$ vertex is formed as a common vertex of a $D^0$ and a $\pi^+$ and it is constrained to coincide with the nearest primary vertex. Muon-tagged candidates are also selected using a dedicated multivariate algorithm aimed at suppressing the combinatorial background. Finally, in events with several $D^{*+}$ and $B$ candidates, only one of them is kept randomly.

Since the detection and production asymmetries are expected to depend on the kinematics of the final-state particles, a weighting procedure is necessary to ensure the cancellation of production and detection asymmetries in Eq.~\eqref{eq:DACP}. The distributions of the transverse momentum, azimuthal angle and pseudorapidity of $D^{*+}$ or $D^0$ mesons are weighted between $K^+K^-$ and $\pi^+\pi^-$ modes. It is then checked \emph{a posteriori} that also the distributions of the tagging pion or muon agree after this procedure.

\begin{figure*}[!t]
\centering
\includegraphics[width=0.24\textwidth]{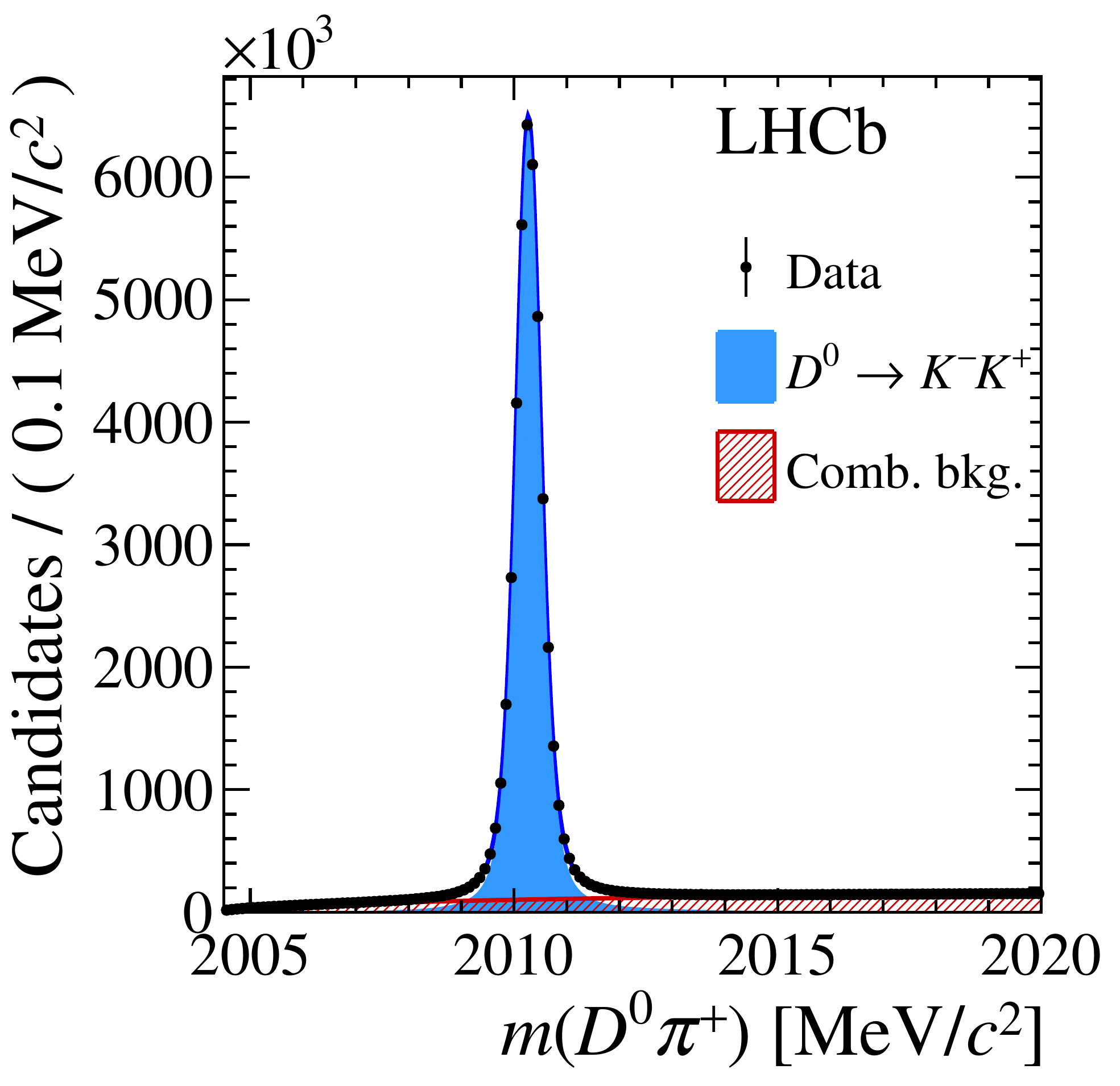}
\includegraphics[width=0.24\textwidth]{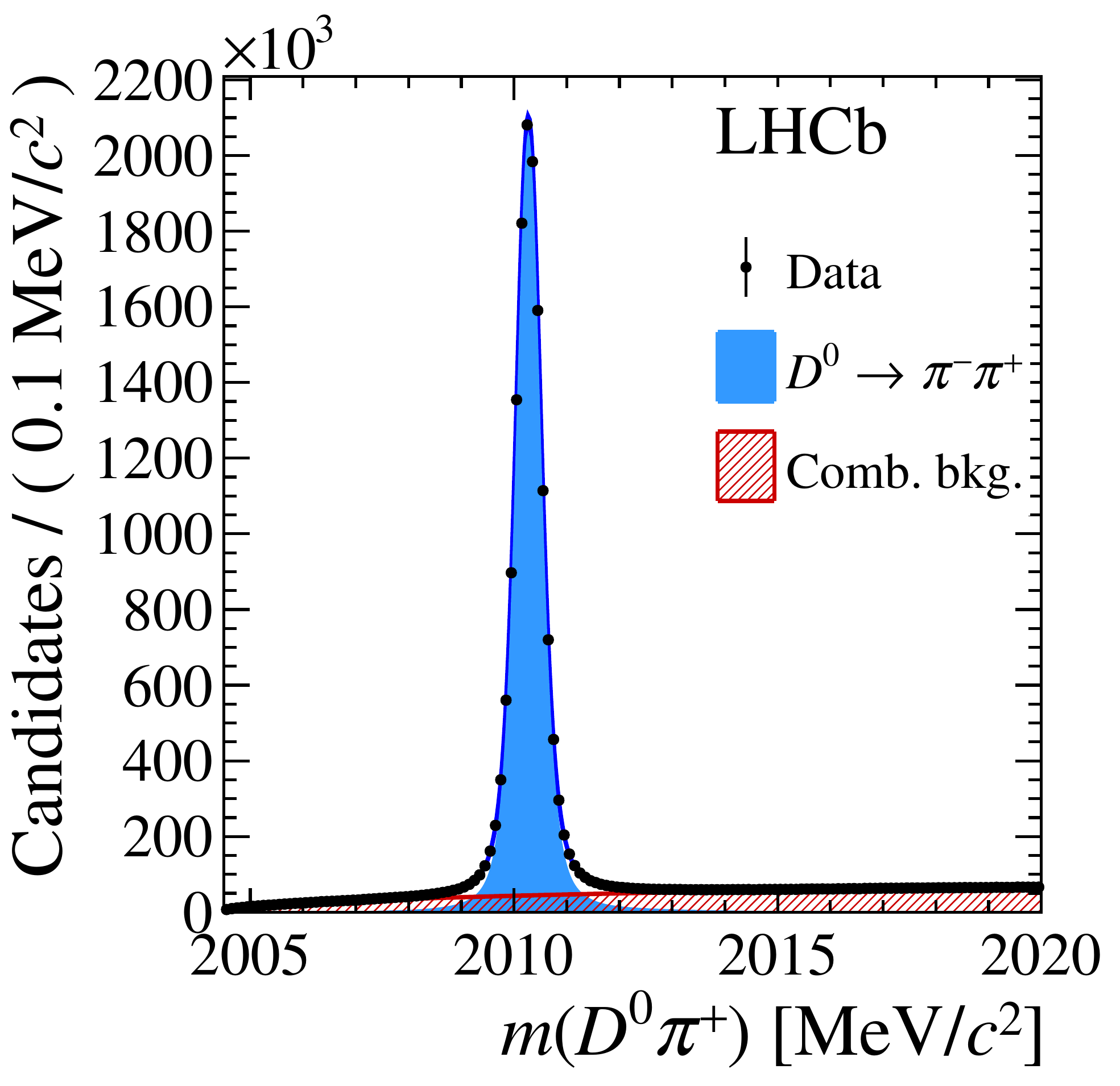}
\includegraphics[width=0.24\textwidth]{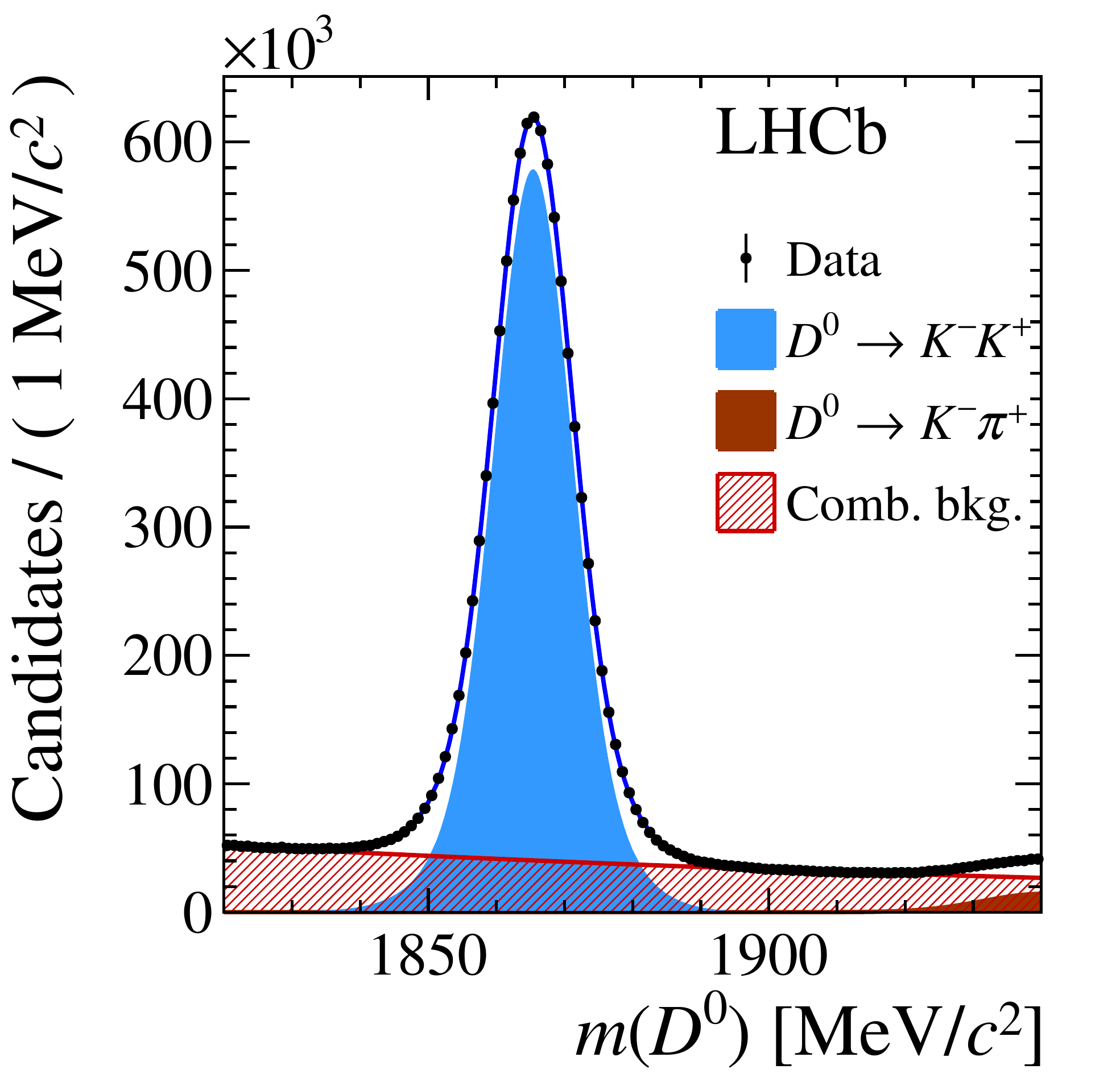}
\includegraphics[width=0.24\textwidth]{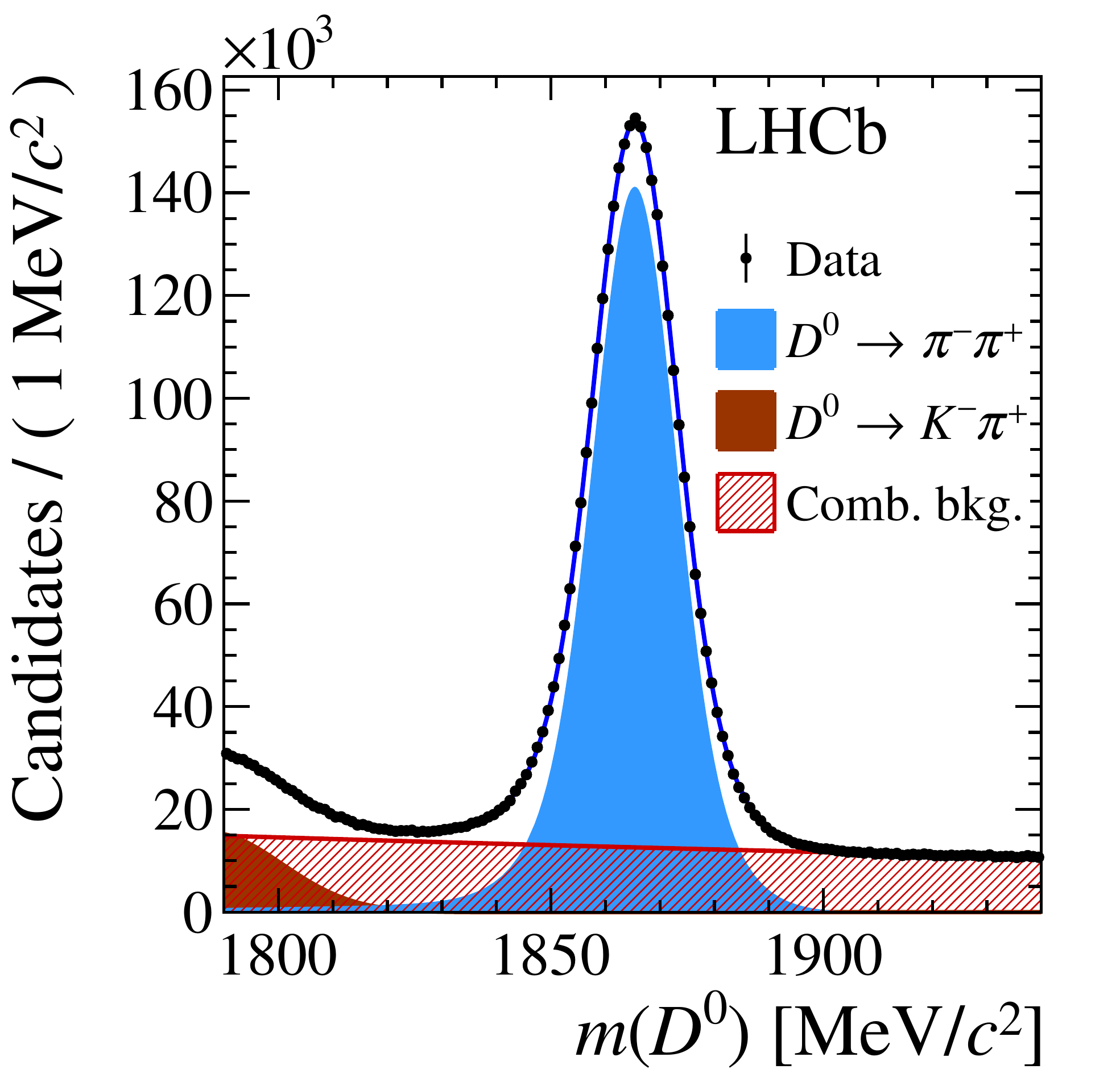}
\caption{Mass distributions of (left) $\pi$-tagged $K^+K^-$, (mid left) $\pi$-tagged $\pi^+\pi^-$, (mid right) $\mu$-tagged $K^+K^-$ and (right) $\mu$-tagged $\pi^+\pi^-$ final states. The fit projections are overlaid.} \label{fig:DACPfit}
\end{figure*}

The raw asymmetries of signal and background components are extracted by means of simultaneous least-squares fits to the binned mass distributions of $D^{*+}$ and $D^{*-}$ $(D^0\ \mathrm{and}\ \overline{D}^0)$ candidates in the prompt (semileptonic) sample. The signal mass model consists of the sum of three Gaussian functions and a Johnson $S_U$ function~\cite{Johnson} in the pion-tagged case, whereas in the muon-tagged case the model is given by the sum of two Gaussian functions convoluted with a truncated power-law function that takes into account final-state photon radiation effects. The combinatorial background is described by an empirical function in the prompt mode and by an exponential function in the semileptonic mode. In the $\mu$-tagged case, the contribution from misidentified $D^0 \to K^-\pi^+$ decays is modeled with the tail of a Gaussian function. The mass distributions with the fit projections overlaid are shown in Fig.~\ref{fig:DACPfit}. The $\pi$-tagged ($\mu$-tagged) signal yields are approximately 44 (9) million \dztokk decays and 14 (3) million \dztopipi decays.

Several sources of systematic uncertainties have been considered. In the prompt case, the main systematic uncertainties are related to the knowledge of the signal and background mass models and to the presence of neglected background components peaking in the $m(D^0)$ distribution. In the semileptonic case, the dominant systematic uncertainty is due to the possibility that the $D^0$ flavor is not determined correctly by the muon charge due to misreconstruction effects.

The results obtained for the difference of the raw asymmetries are
\begin{eqnarray}
\DACPprompt &=& (-18.2 \pm 3.2 \pm 0.9) \times 10^{-4}, \nonumber \\
\DACPSL &=& (-9 \pm 8 \pm 5) \times 10^{-4}, \nonumber
\end{eqnarray}
where the first uncertainties are statistical and the second systematic.
By combining these results with previous LHCb measurements~\cite{Aaij:2014gsa,Aaij:2016cfh} the following value is obtained
\begin{equation}
\DACP = (-15.4 \pm 2.9)\times 10^{-4}, \nonumber
\end{equation}
where the uncertainty contains both statistical and systematic contributions. The significance of the deviation from zero is 5.3$\sigma$. This is the first observation of \CP violation in the decay of charm hadrons.

By using in Eq.~\eqref{eq:DACP} the values of $\Delta \langle t \rangle / \tau(D^0) = 0.115 \pm 0.002$ measured from the dataset and obtained using the world average of the $D^0$ lifetime~\cite{PDG2018} and the LHCb average $A_\Gamma = (-2.8\pm2.8)\times 10^{-4}$~\cite{Aaij:2015yda,Aaij:2017idz} it is possible to obtain
\begin{equation}
\Delta a^{\mathrm{dir}}_{\CP} = (-15.7 \pm 2.9) \times 10^{-4}, \nonumber
\end{equation}
which confirms that $\Delta A_{\CP}$ is mainly sensitive to direct \CP violation.

\section{Search for \boldmath{\CP} violation in \boldmath{$D_s^+ \to K^0_s \pi^+$, $D^+ \to K^0_s K^+$} and \boldmath{$D^+ \to \phi \pi^+$ decays}} \label{sec:ACPDp}
In these decays, \CP violation can arise in the interference between loop- and tree-level processes in the Cabibbo suppressed $c \to d\overline{d}u$ and $c \to s\overline{s}u$ transitions. The search for \CP violation in these channels is important since beyond Standard Model (SM) processes could significantly enhance the size of \CP violation expected in these decays~\cite{Grossman:2006jg}.

The raw asymmetry, defined as the difference between the signal yields of positively and negatively charged $D^{+}_{(s)}$ mesons normalised to the sum for each mode, can be written as
\begin{equation} \label{eq:ArawD2KS0h}
A_\mathrm{raw}^{D^{+}_{(s)} \to f^+} \approx A_{\CP}^{D^{+}_{(s)} \to f^+} + A_\mathrm{P}^{D^{+}_{(s)}} + A_\mathrm{D}^{f^+},
\end{equation}
where the first term on the right-hand side is the \CP asymmetry of the considered mode, the second is the $D^{+}_{(s)}$ production asymmetry and the last is the $f^+$ detection asymmetry, with $f^+ = K^0_s\pi^+,\ K^0_s K^+$ or $\phi \pi^+$.
The detection and production asymmetries are canceled using samples of Cabibbo-favoured decays, such as $D^+ \to K^0_s \pi^+$, $D^+_s \to K^0_s K^+$ and $D^+_s \to \phi \pi^+$, for which the \CP asymmetries are expected to be negligibly small compared to the Cabibbo-suppressed modes. The \CP asymmetries for the decay modes of interest are then determined as
\begin{eqnarray}
\ACPDspKspi &\approx& \ArawDspKspi - \ArawDspphipi , \label{eq:ACPDspKspi} \\
\ACPDpKsK &\approx& \ArawDpKsK - \ArawDpKspi  \nonumber \\
 &-& \ArawDspKsK + \ArawDspphipi, \label{eq:ACPDpKsK}\\
\ACPDpphipi &\approx& \ArawDpphipi - \ArawDpKspi, \label{eq:ACPDpphipi}
\end{eqnarray}
where the contribution from the $K^0$ detection asymmetry has been omitted and must be subtracted where relevant.

The dataset used corresponds to an integrated luminosity of 3.8\invfb collected by the LHCb collaboration during part of 2015-2017.
The data are selected requiring that one or more of the $D^{+}_{(s)}$ decay products are associated with a large transverse energy deposit in the LHCb calorimeter system and by two levels of software trigger. The $K^0_s$ candidates are built from pions that are reconstruced in the vertex detector, to reduce to a negligible level the interference between Cabibbo-favoured and doubly Cabibbo-suppressed amplitudes resulting from kaon mixing. Finally, the candidates with a $K^0_s$ in the final state are further selected by means of an artificial neural network using both kinematic and geometric quantities. The $\phi(1020)$-meson mass is required to be within 10 MeV/$c^2$ from its PDG value~\cite{PDG2018}. Since the contribution of $D^{+}_{(s)}$ mesons produced from $b$-hadron decays could bias the measurement, a cut on the impact parameter in the plane transverse to the beam (TIP) is necessary to reduce this effect. The residual contribution from secondary decays is accounted for with a systematic uncertainty. The backgrounds due to the misidentification of one of the final-state particles are reduced to negligible levels using PID requirements and kinematic vetoes.
Fiducial requirements are used to exclude the regions with large detection asymmetries.

Since the detection and production asymmetries could depend on the kinematics of the involved particles, a weighting procedure is necessary in order to ensure an adequate cancellation of the nuisance asymmetries.

In the measurement of $D^{+}_{s} \to K^0_s \pi^+$ $(D^+ \to \phi \pi^+)$ \CP asymmetries, the kinematic distributions of $D^{+}_s$ and $\pi^+$ ($D^{+}$ and $\pi^+$) mesons are weighted between the relevant modes. For the measurement of the $D^+ \to K^0_s K^+$ \CP asymmetry, the kinematic distributions of $D^+$, $K^+$ and $\pi^+$ are weighted between all relevant modes.

\begin{figure*}[!t]
\centering
\includegraphics[width=0.32\textwidth]{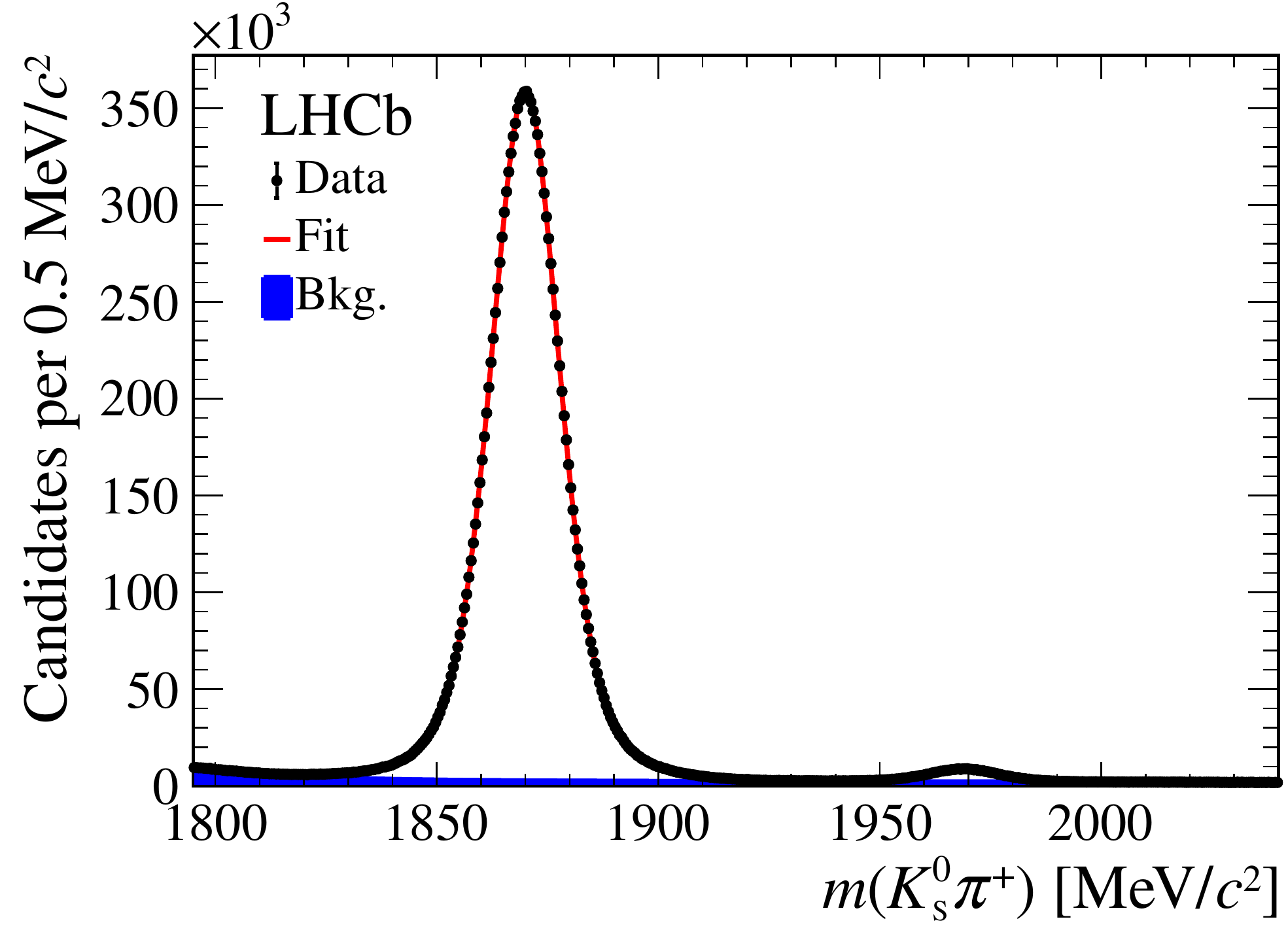}
\includegraphics[width=0.32\textwidth]{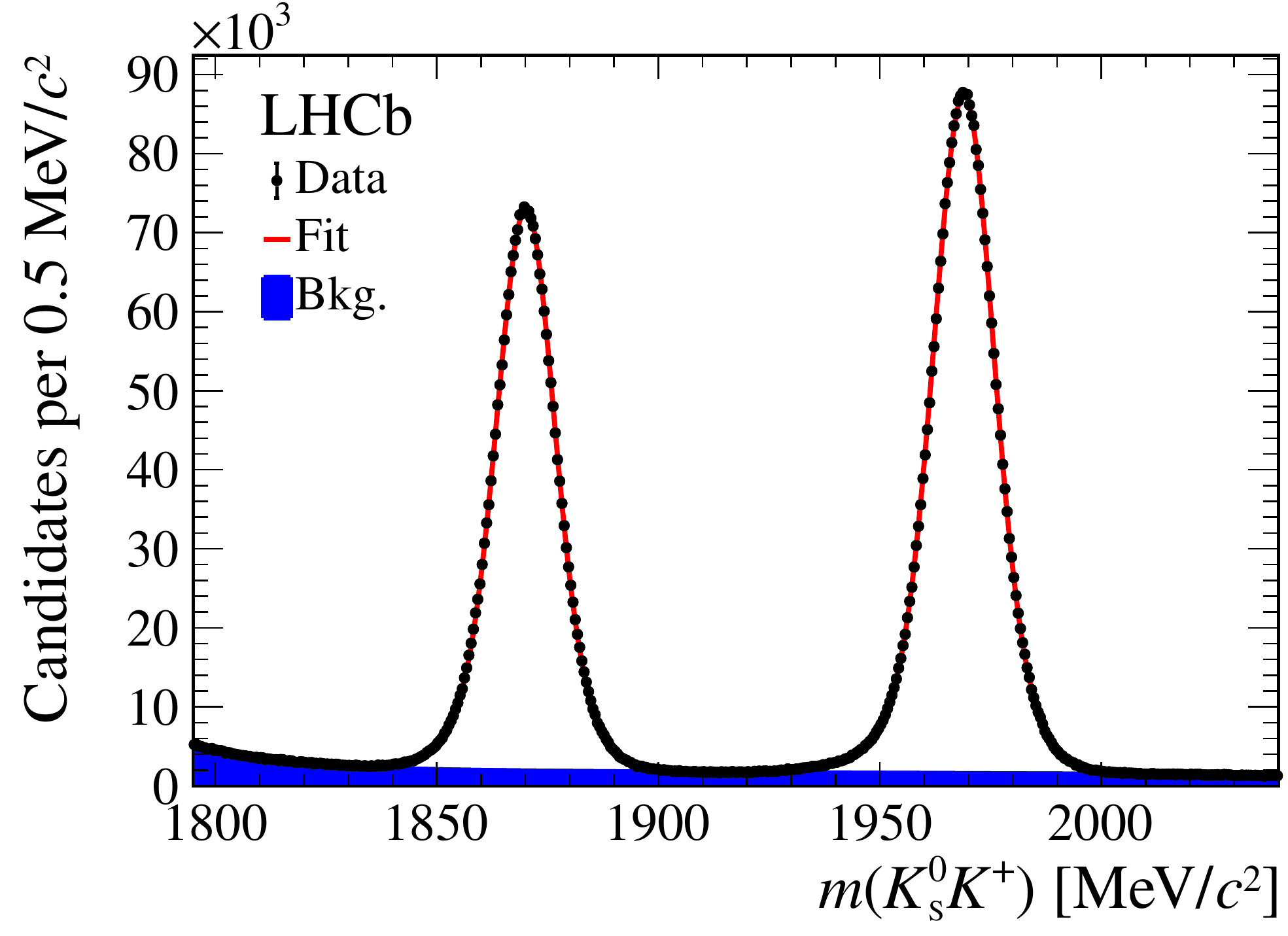}
\includegraphics[width=0.32\textwidth]{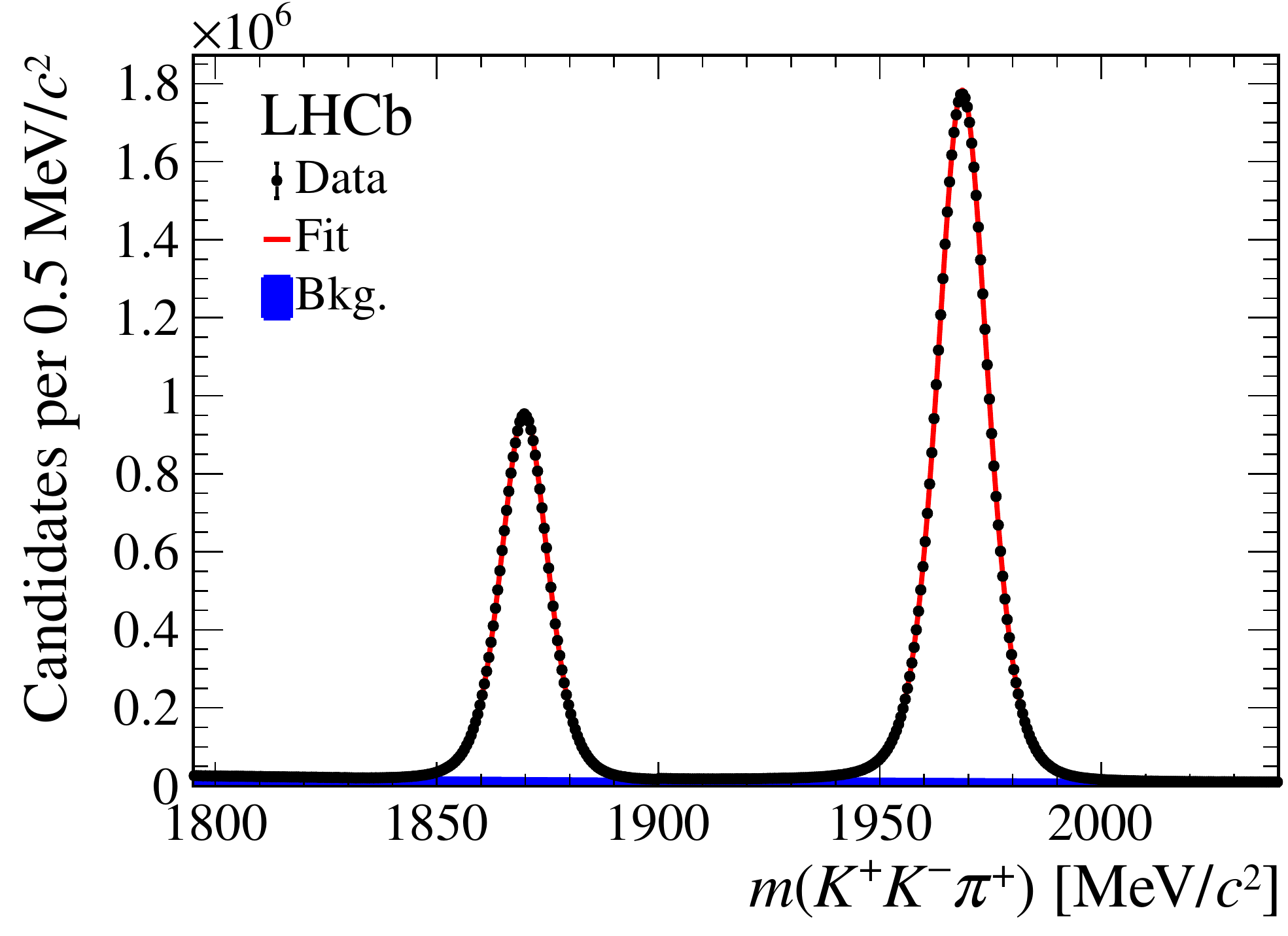}
\caption{Mass distributions of (left) $D^{+}_{s} \to K^0_s \pi^+$, (middle) $D^{+}_{s} \to K^0_s K^+$ and (right) $D^{+}_{s} \to \phi \pi^+$ candidates. The fit projections are overlaid.} \label{fig:ACPKshfit}
\end{figure*}

The raw asymmetries of all the modes entering Eqs.~\ref{eq:ACPDspKspi},~\ref{eq:ACPDpKsK} and~\ref{eq:ACPDpphipi} are determined by means of least-squares fits to the invariant-mass distributions of $D^{+}_{s}$ and $D^{-}_{s}$ candidates. In each fit the signal component is modeled as the sum of a Gaussian function with a Johnson $S_U$ function~\cite{Johnson}, whereas the combinatorial background is described by the sum of two exponential functions. The mass distributions with the fit projections are shown in Fig.~\ref{fig:ACPKshfit}.

Several sources of systematic uncertainties have been considered. The dominant systematic uncertainty for all the modes is due to the assumed signal and background shapes used in the fit. This is evaluated by fitting with the default model large sets of pseudoexperiments where alternative models that describe data equally well are used in generation.

After correcting for the $K^0$ detection asymmetry by following the technique developed in Ref.~\cite{Aaij:2014gsa}, the following \CP asymmetries are obtained
\begin{eqnarray}
\ACPDspKspi &=& (\phantom{-} 1.3 \pm 1.9 \pm 0.5) \times 10^{-3}, \nonumber \\
\ACPDpKsK  &=& (-0.09 \pm 0.65 \pm 0.48) \times 10^{-3}, \nonumber \\
\ACPDpphipi &=& (\phantom{-} 0.05 \pm 0.42 \pm 0.29) \times 10^{-3}, \nonumber
\end{eqnarray}
where the first uncertainties are statistical and the second systematic.
These results represent the best determination of these quantities to date and are consistent with the hypothesis of no \CP violation.

\section{Search for \boldmath{\CP} violation with kinematic asymmetries in the \boldmath{$D^0 \to K^+ K^- \pi^+ \pi^-$} decay}
Charge-parity violation in Cabibbo-suppressed $D^0 \to K^+ K^- \pi^+ \pi^-$ decays has been studied by means of $\hat{T}$-odd correlations~\cite{Link:2004wx,delAmoSanchez:2010xj,Aaij:2014qwa}, where $\hat{T}$ stands for the operator that reverses the direction of momenta and spins. Amplitudes of the decay can be extracted from $A_X$, which it is defined as
\begin{equation}
A_X \equiv \frac{\Gamma(X>0) - \Gamma(X<0)}{\Gamma(X>0) + \Gamma(X<0)}
\end{equation}
where $X$ is a kinematic variable and $\Gamma$ is the decay rate of $D^0$ decays. The \CP-violating kinematic asymmetry can then be written as
\begin{equation}
a_{\CP}^X \equiv \frac{1}{2} \left( A_X - \eta_{\CP}^X \overline{A}_{\overline{X}}     \right)
\end{equation}
where $\eta_{\CP}^X$ is a \CP eigenvalue specific of X. The chosen kinematic variables are combinations of $\theta_1$, $\theta_2$ and $\Phi$. The $\theta_1$ $(\theta_2)$ variable is the angle between the $K^+$ $(\pi^+)$ meson and the direction opposite to that of the $D^0$ momentum in the centre-of-mass frame of the $K^+K^-$ $(\pi^+\pi^-)$ system. The $\Phi$ variable is the angle between the $K^+K^-$ and $\pi^+\pi^-$ pairs in the centre-of-mass frame. A set of five kinematic variables is defined: $\sin 2\Phi$, $\cos \theta_1 \cos \theta_2 \sin \Phi$, $\sin \Phi$, $\cos \Phi$ and $\cos \theta_1 \cos \theta_2 \cos \Phi$. The first three variables have $\eta_{\CP}^X = -1$, whereas the others have $\eta_{\CP}^X = +1$.

In this analysis the total dataset collected by the Belle detector, corresponding to an integrated luminosity of 988\invfb, is used. The flavor of the $D^0$ is identified by the charge of the accompanying pion in the $D^{*+} \to D^0 \pi^+$ strong decay. The final-state particles are selected imposing requirements on their momenta and transverse momenta. The charged tracks are identified as pions or kaons depending on the ratio of their PID likelihoods. Vetoes are applied in order to reject $D^0 \to K^+ K^- K^0_s$ decays that could fake a signal. Finally, in events with more than one candidate, only one candidate is retained.

\begin{figure}[!t]
\centering
\includegraphics[width=0.23\textwidth]{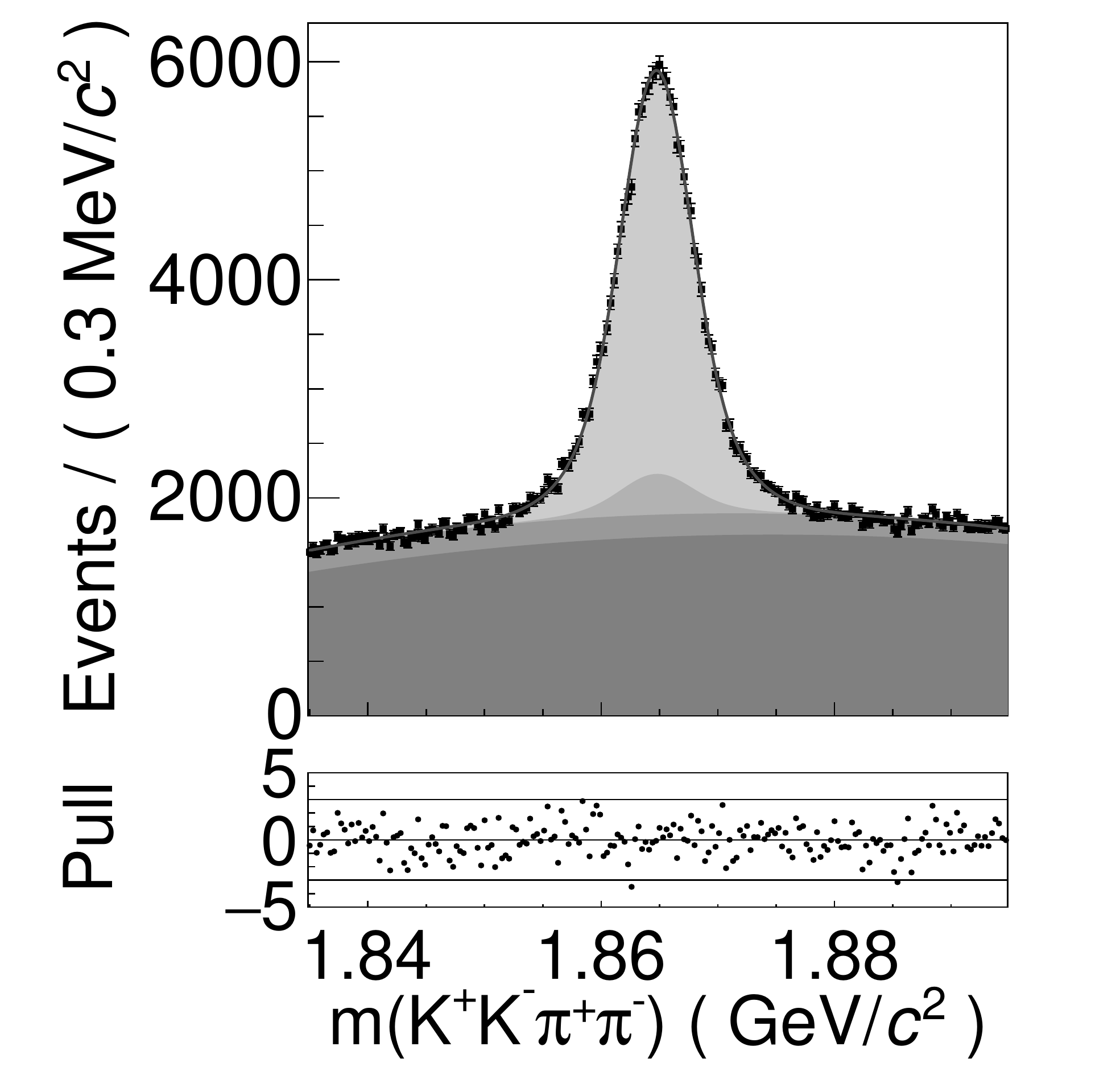}
\includegraphics[width=0.23\textwidth]{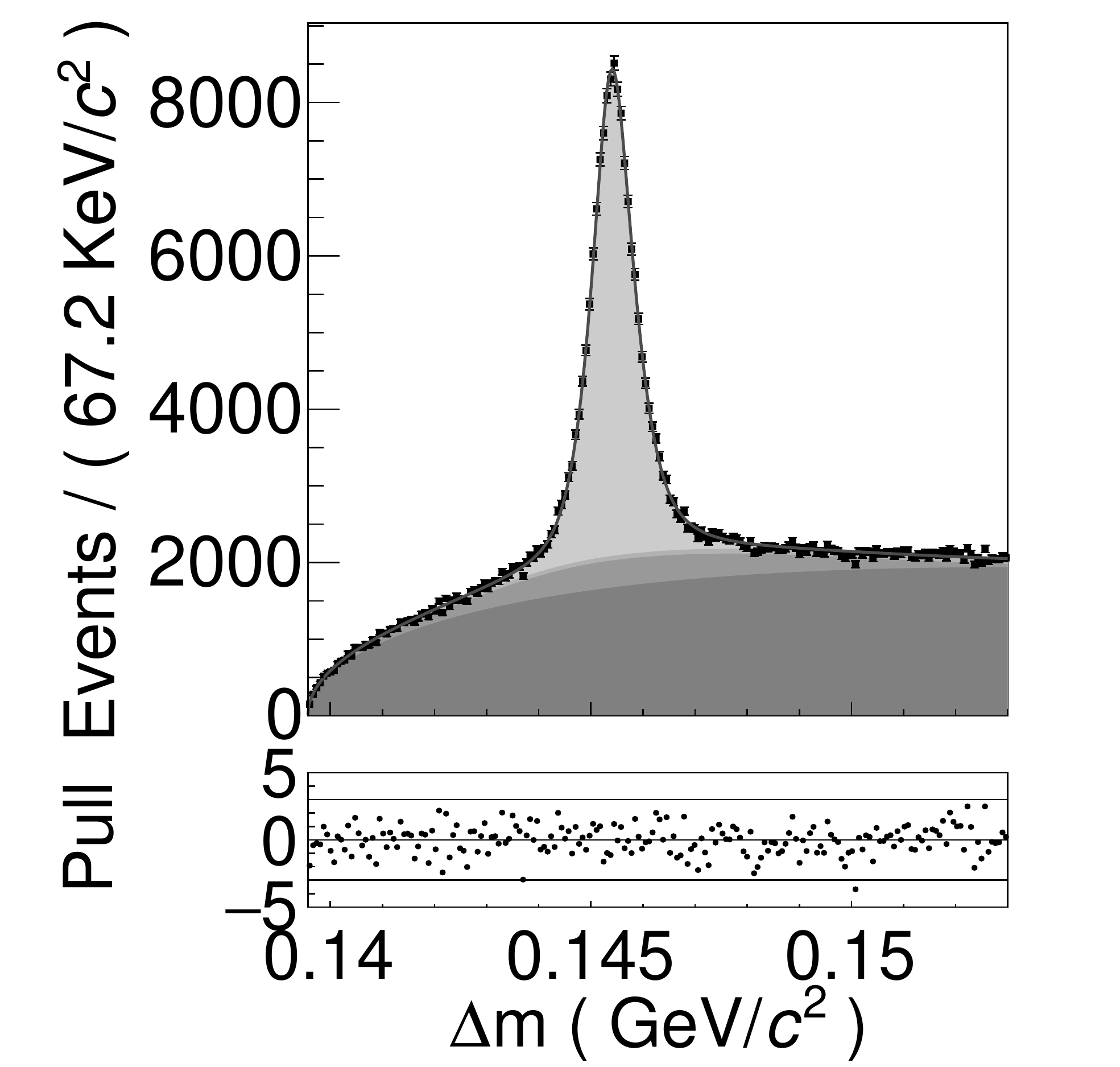}
\caption{Distributions of (left) $m(K^+ K^- \pi^+ \pi^-)$ and (right) $\Delta m$ with the results of the fit overlaid. The different components, from lowest to highest are: combinatorial background, partially reconstructed $D^0$ background, random-pion background and signal.} \label{fig:fitKinAsym}
\end{figure}

The \CP-violating asymmetries are calculated by comparing the signal yields for each $D^0$ flavor and sign of the kinematic variable. Binned two-dimensional maximum-likelihood fits are performed to the $m(K^+ K^- \pi^+ \pi^-)$ and $\Delta m = m(D^0\pi^+) - m(D^0)$ distributions. The signal component is modeled as the sum of a Gaussian and a bifurcated Gaussian function for $m(K^+ K^- \pi^+ \pi^-)$ and as the sum of a Gaussian function and a Johnson $S_U$ function for $\Delta m$. The combinatorial background component is described by a Chebyshev function for $m(K^+ K^- \pi^+ \pi^-)$ and a threshold function for $\Delta m$. The background due to the association of a $D^0$ with a random pion is the product of the signal shape for $m(K^+ K^- \pi^+ \pi^-)$ and of the combinatorial shape for $\Delta m$. Finally, the background due to partially reconstructed $D^0$ mesons is described by a Chebyshev function for $m(K^+ K^- \pi^+ \pi^-)$ and a bifurcated Gaussian function for $\Delta m$. The $m(K^+ K^- \pi^+ \pi^-)$ and $\Delta m$ distributions with the fit projections overlaid are shown in Fig.~\ref{fig:fitKinAsym}.

Several sources of systematic uncertainty are considered. The dominant one is due to the detector bias and it is estimated with a control sample of Cabibbo-favoured $D^0 \to K^+ \pi^- \pi^+ \pi^-$ decays in which all the kinematic asymmetries are expected to be well below the experimental precision. The measurement of each kinematic asymmetry is repeated on this sample and the associated statistical uncertainty is taken as systematic uncertainty for the corresponding quantity.

The \CP-violating kinematic asymmetries are found to be
\begin{eqnarray}
a^{\CP}_{\cos \Phi} &=& (\phantom{-}3.4 \pm 3.6 \pm 0.6) \times 10^{-3}, \nonumber \\
a^{\CP}_{\sin \Phi} &=& (\phantom{-}5.2 \pm 3.7 \pm 0.7) \times 10^{-3}, \nonumber \\
a^{\CP}_{\sin 2\Phi} &=& (\phantom{-}3.9 \pm 3.6 \pm 0.7) \times 10^{-3}, \nonumber \\
a^{\CP}_{\cos \theta_1 \cos \theta_2 \cos \Phi} &=& (-0.2 \pm 3.6 \pm 0.7) \times 10^{-3}, \nonumber \\
a^{\CP}_{\cos \theta_1 \cos \theta_2 \sin \Phi} &=& (\phantom{-}0.2 \pm 3.7 \pm 0.7) \times 10^{-3}, \nonumber
\end{eqnarray}
where the first uncertainties are statistical and the second systematic. No \CP violation is observed within the current experimental precision.

\section{Search for time-dependent \boldmath{\CP} violation in \boldmath{$D^0 \to K^+ K^-$} and \boldmath{$D^0 \to \pi^+ \pi^-$} decays}
Tests of \CP violation in the mixing or interference between mixing and decay, for which the SM predictions still lie about one order of magnitude below the current experimental precision~\cite{Cerri:2018ypt}, are complementary to those of direct \CP violation and might help clarify the picture after the first observation of \CP violation in charm decays. Since the charm mixing parameters are both smaller than $10^{-2}$~\cite{Amhis:2016xyh}, the time-dependent \CP asymmetry, defined as the difference of the decay rates of $D^0$ and $\overline{D}^0$ mesons to a final state $f=K^+K^-$ or $\pi^+\pi^-$ normalized to the sum of the decay rates, can be approximated as
\begin{equation}
A_{\CP}(f,t) \approx \Adir  - A_{\Gamma}(f) \frac{t}{\tau_{D^0}},
\end{equation}
that is the equivalent of Eq.~\eqref{eq:ACP} without the averaging on the decay time.

The flavor of the $D^0$ at the production is inferred from the charge of the accompanying pion in $D^{*+} \to D^0 \pi^+$ strong decays.
The time-dependent raw asymmetry can be calculated as the difference between positively and negatively tagged candidates normalized to their sum and can be exapanded as
\begin{equation}
A_{\mathrm{raw}} (f,t) \approx A_{\CP}(f,t) + A_\mathrm{D}(\pi) + A_{\mathrm{P}}(D^{*}),
\end{equation}
where $A_\mathrm{D}(\pi)$ and $A_{\mathrm{P}}(D^{*})$ are the pion detection and the $D^{*}$ production asymmetries, respectively. Once the raw asymmetry is measured as a function of $D^0$ decay time, $A_\Gamma$ can be obtained from a fit with a linear function to the $A_{\mathrm{raw}}(f,t)$ values.

The dataset used corresponds to an integrated luminosity of about 2\invfb collected by the LHCb experiment during 2015 and 2016. The data are selected by a two-stage software trigger, without imposing any requirement on the output of the hardware trigger. The first stage of the software trigger requires that at least one track has high transverse momentum and impact parameter. At the second stage, two oppositely-charged and high quality tracks with momenta above 5 GeV/$c$ and a distance of closest approach less than 0.1 mm are combined to form a $D^0$-meson candidate. Finally, a high-quality pion with momentum above 1 GeV/$c$ and transverse momentum above 100 MeV/$c$ is combined with the $D^0$ meson to form a $D^{*+}$ candidate. Particle identification criteria are applied offline in order to select $K^+K^-$, $\pi^+\pi^-$ and $K^- \pi^+$ final states. Moreover, the $D^0$ meson is required to have a flight distance along the $z$ axis (in the $x-y$ plane) greater than 3 mm (0.3 mm) in order to reject large detection asymmetry regions. Finally, in events with more than one $D^*$ candidate, only one of them is kept on a random basis.

\begin{figure}[!t]
\centering
\includegraphics[width=0.33\textwidth]{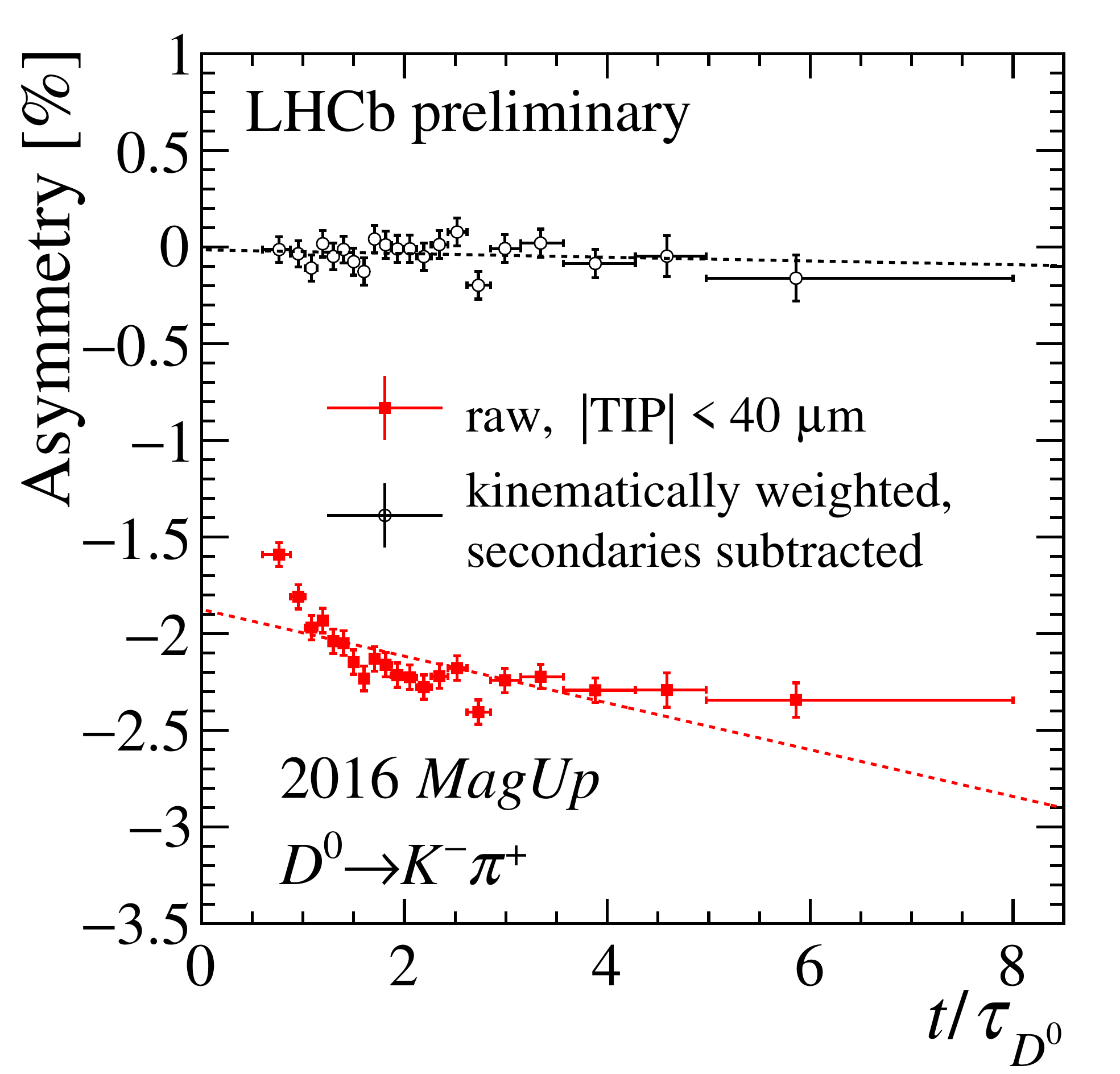}
\caption{Values of the raw asymmetry for the $K^-\pi^+$ control sample and 2016 \emph{MagUp} dataset (red full circles) before and (black empty circles) after the weighting procedure. The results of a fit with a linear function are shown as dashed lines in both cases.} \label{fig:ArawAGamma}
\end{figure}

The data sample contains momentum-dependent charge asymmetries due to the pion used to tag the $D^0$ flavor (as the ones described in Secs.~\ref{sec:DACP} and~\ref{sec:ACPDp}) that could bias the measurement. This effect is corrected for by weighting the $(q_\mathrm{tag}\theta_x,\theta_y,k)$ distributions of $D^0$ and $\overline{D}^0$ mesons to their average value, where $q_\mathrm{tag}$ is the charge of the tagging pion, $\theta_{x(y)} \equiv \arctan(p_{x(y)}/p_z)$ and $k \equiv 1/\sqrt{p_x^2 + p_y^2}$. The effect of this correction (shown in Fig.~\ref{fig:ArawAGamma}) and the whole analysis procedure are checked on a large sample of Cabibbo-favoured $D^0 \to K^-\pi^+$ decays, for which the value of $A_\Gamma$ is expected to be well below the experimental precision. The pollution from secondary decays is accounted for with a fit to the TIP distribution in each decay-time bin. The prompt component is modeled with a Dirac delta centered at the value TIP = 0 convoluted with a Gaussian resolution function, while the secondary component is described empirically with an exponential function convoluted with a Gaussian resolution function. In order to measure $A_\Gamma$, a linear function is fitted to the time-dependent asymmetry of primary decays obtained through fits to the TIP distribution in 21 bins of decay time in the range $[0.6,8]\tau_{D^0}$.

Several sources of systematic uncertainties are evaluated. The dominant ones are due to the secondary decays contamination and to the removal of the background under the $\Delta m = m(h^+h^-\pi^+) - m(h^+h^-)$ peak. The first uncertainty is evaluated by using different resolution functions for the TIP, as well as other parameterisations for the secondary component. The second uncertainty is estimated by changing the definition of the sideband used to perform the background-subtraction of the $\Delta m$ distribution.

\begin{figure}[!t]
\centering
\includegraphics[width=0.4\textwidth]{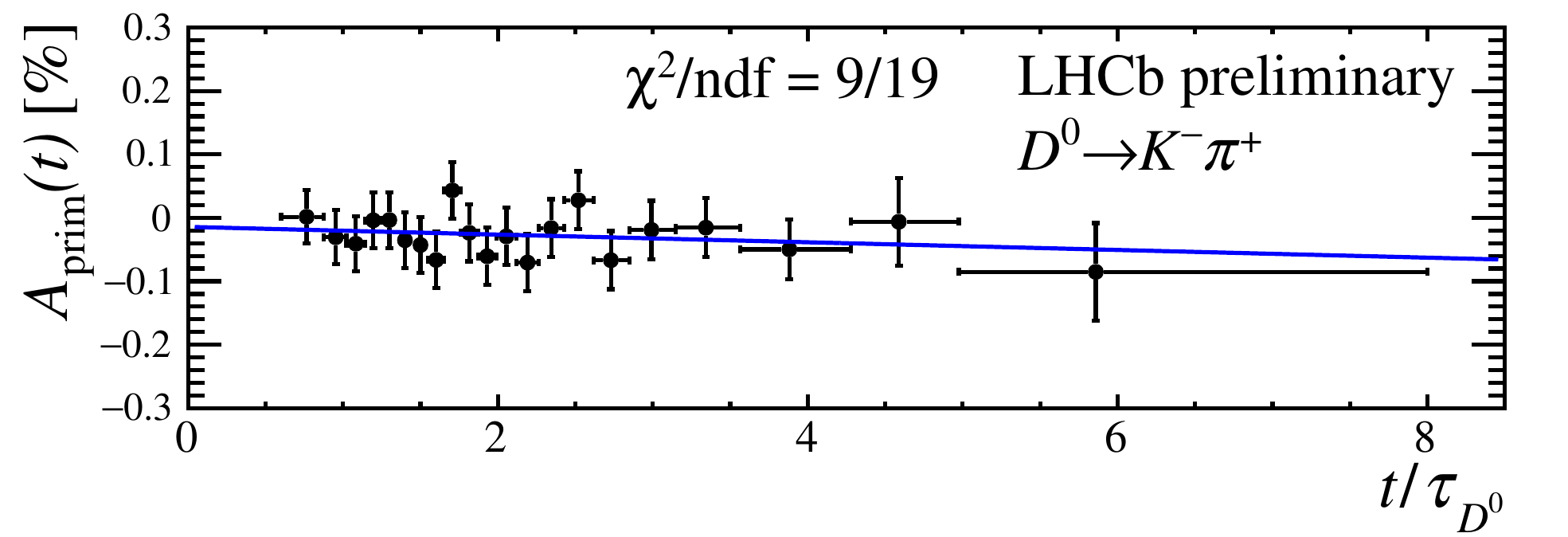}
\includegraphics[width=0.4\textwidth]{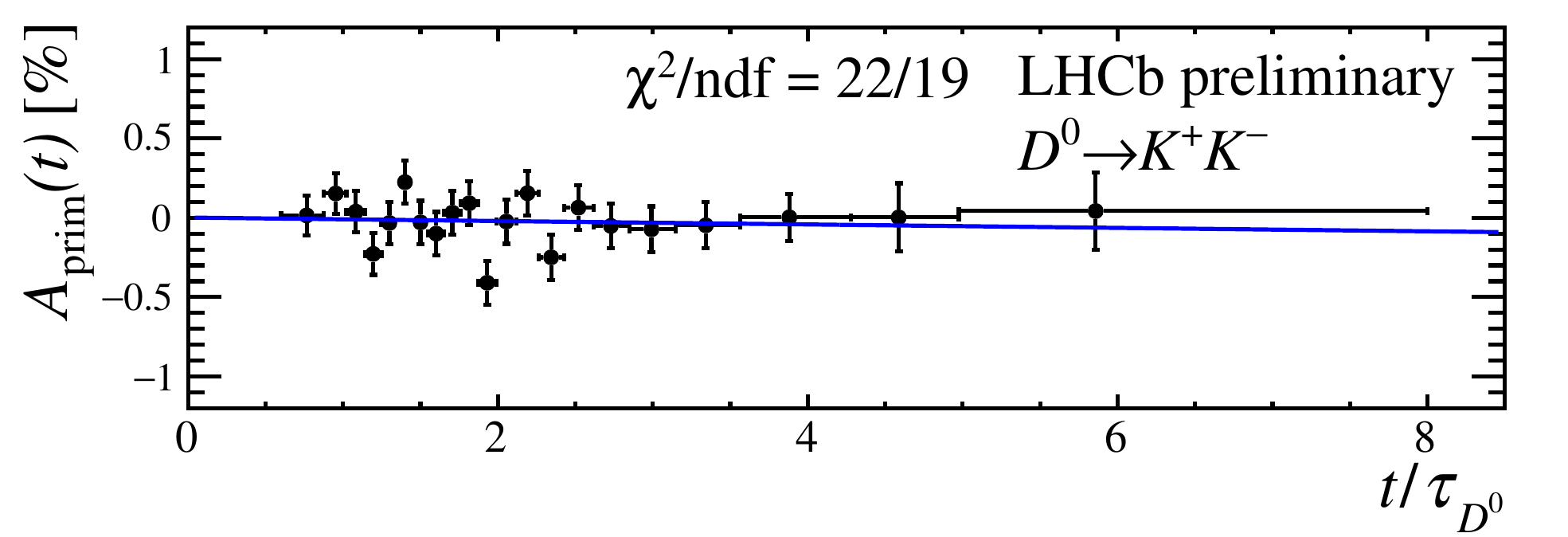}
\includegraphics[width=0.4\textwidth]{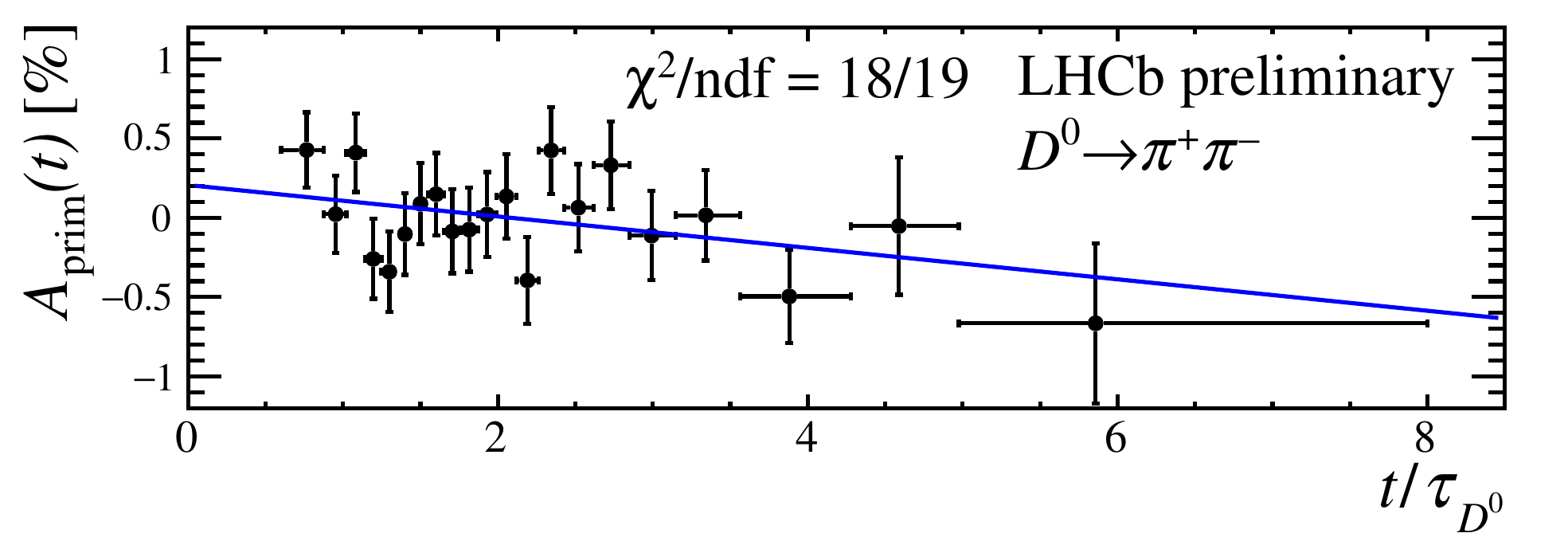}
\caption{Measured values of the time-dependent asymmetry for the primary component for (top) $D^0 \to K^- \pi^+$, (middle) $D^0 \to K^+ K^-$ and (bottom) $D^0 \to \pi^- \pi^+$ decays. The fit projections are overlaid.} \label{fig:fitAgamma}
\end{figure}

The result obtained on the control sample of Cabibbo-favoured $D^0 \to K^- \pi^+$ decays is
\begin{equation}
A_\Gamma(K^- \pi^+) = (0.7 \pm 1.1)\times 10^{-4}, \nonumber
\end{equation}
where the uncertainty is only statistical.
The results obtained for the $K^+K^-$ and $\pi^+\pi^-$ modes are
\begin{eqnarray}
A_\Gamma(K^+ K^-) &=& (1.3 \pm 3.5 \pm 0.7)\times 10^{-4}, \nonumber \\
A_\Gamma(\pi^+ \pi^-) &=& (11.3 \pm 6.9 \pm 0.8)\times 10^{-4}, \nonumber
\end{eqnarray}
where the first uncertainties are statistical and the second systematic. The projections of the fits to the time-dependent asymmetries are shown in Fig.~\ref{fig:fitAgamma}. If weak phases in the decay are neglected~\cite{Grossman:2006jg,Du:2006jc}, $A_\Gamma$ does not depend on the $D^0$ decay channel and the two values can be combined. The average value of $A_\Gamma$ is
\begin{equation}
A_\Gamma (K^+K^- + \pi^+\pi^-) = (3.4 \pm 3.1 \pm 0.6)\times 10^{-4}, \nonumber
\end{equation}
where the first uncertainty is statistical and the second systematic. This value is consistent with the hypothesis of no \CP violation in mixing or in the interference between mixing and decay.

\section{Measurement of the mass difference between neutral charm-meson eigenstates}
Direct experimental access to charm-mixing parameters is offered by self-conjugate multibody decays, such as $D^0 \to K^0_s \pi^+ \pi^-$. For neutral charm mesons it is possible to write $| D_{1,2} \rangle \equiv p |D^0 \rangle + q | \overline{D}^0 \rangle$, where $q$ and $p$ are complex parameters. If $|q/p| \neq 1$ then \CP symmetry is violated in mixing, while if $\phi_f \equiv \arg (q\overline{A}_f/pA_f) \neq 0$ then \CP symmetry is violated in the interference between mixing and decay. The amplitude $A_f(\overline{A}_f)$ refers to the decay $D^0 \to K^0_s \pi^+ \pi^-(\overline{D}^0 \to K^0_s \pi^+ \pi^-)$. If \CP symmetry is conserved in the decay amplitude $(|A_f|^2 = |\overline{A}_f|^2)$, then $\phi_f$ is independent of the final state, $\phi_f \approx \phi = \arg(q/p)$.
One can then define the dimensionless parameters $x \equiv (m_1-m_2)/\Gamma$ and $y \equiv (\Gamma_1 - \Gamma_2)/(2\Gamma)$, where $m_{1(2)}$ and $\Gamma_{1(2)}$ are the mass and decay width of the $D_{1(2)}$ mass eigenstate, and $\Gamma = (\Gamma_1 + \Gamma_2)/2$~\cite{PDG2018}. In this analysis, a novel model-independent approach, called the bin-flip method, is adopted. This approach is optimized for the measurement of the parameter $x$~\cite{DiCanto:2018tsd}.

The Dalitz plot is divided in different bins that preserve nearly constant strong-phase differences, $\Delta \delta(m^2_{-},m^2_{+})$, where $m^2_{\pm}$ means $m^2(K^0_s\pi^{\pm})$ for $D^0 \to K^0_s \pi^+ \pi^-$ and $m^2(K^0_s \pi^{\mp})$ for $\overline{D}^0 \to K^0_s \pi^+ \pi^-$ decays. Two sets of eight bins are formed symmetrically about the bisector $m^2_{+} = m^2_{-}$. Positive indices refer to bins where $m^2_{+} > m^2_{-}$, whereas negative indices refer to bins where $m^2_{+} < m^2_{-}$. The data are further split in bins of decay time, labelled with the $j$ index. For each bin, the ratio $R^{+}_{bj}(R^{-}_{bj})$ between initially produced $D^0(\overline{D}^0)$ mesons in Dalitz bin $-b$ and Dalitz bin $b$ is measured. The ratios are proportional to a combination of terms involving the parameter $z_{\CP} \pm \Delta z \equiv -(q/p)^{\pm1}(y + ix)$ and to the strong-phase differences taken as external inputs~\cite{DiCanto:2018tsd}. The results are expressed in terms of the \CP-averaged mixing parameters $x_{\CP} \equiv -\Im(z_{\CP})$, $y_{\CP} \equiv - \Re(z_{\CP})$, $\Delta x \equiv - \Im(\Delta z)$ and $\Delta y \equiv - \Re(\Delta z)$.

The dataset used corresponds to an integrated luminosity of 3\invfb collected by the LHCb experiment during 2011 and 2012. Both prompt $D^{*+} \to D^0 \pi^+$ and semileptonic $\overline{B} \to D^0 \mu^- X$ decays are used.
The selection applies criteria consistent with the decay topology on momenta, vertex and track displacements, PID information and invariant masses of either $D^{*+}$ or $D^0$ decay products, depending on the considered sample. One random candidate is retained in events where multiple signal candidates are present.

The signal yields in each bin are obtained by fits to the $\Delta m \equiv m(K^0_s \pi^+ \pi^- \pi^+) -  m(K^0_s \pi^+ \pi^-)$ distribution for the prompt sample. In the semileptonic case, the signal yields are obtained by fits to the $m(K^0_s \pi^+ \pi^-)$ distribution. The mixing parameters are obtained by minimizing a least-square function that compares the decay-time evolution of signal yields observed in Dalitz bins $- b$ and $+ b$, along with their uncertainties.

The dominant systematic uncertainty on $x_{\CP}$ is due to the contamination of secondary decays in the prompt sample and to the presence of $D^0$ mesons combined with random muons in the semileptonic sample. The main systematic uncertainty on $y_{\CP}$ is due to the neglected decay time and $m^2_{\pm}$ resolutions, as well as the neglected efficiency variations across the Dalitz plane. Finally, possibly nonuniformities with respect to the two halves of the Dalitz plot induced by reconstruction inefficiencies are the main systematic uncertainty in the determination of $\Delta x$ and $\Delta y$.

\begin{figure}[!tb]
\centering
\includegraphics[width=0.36\textwidth]{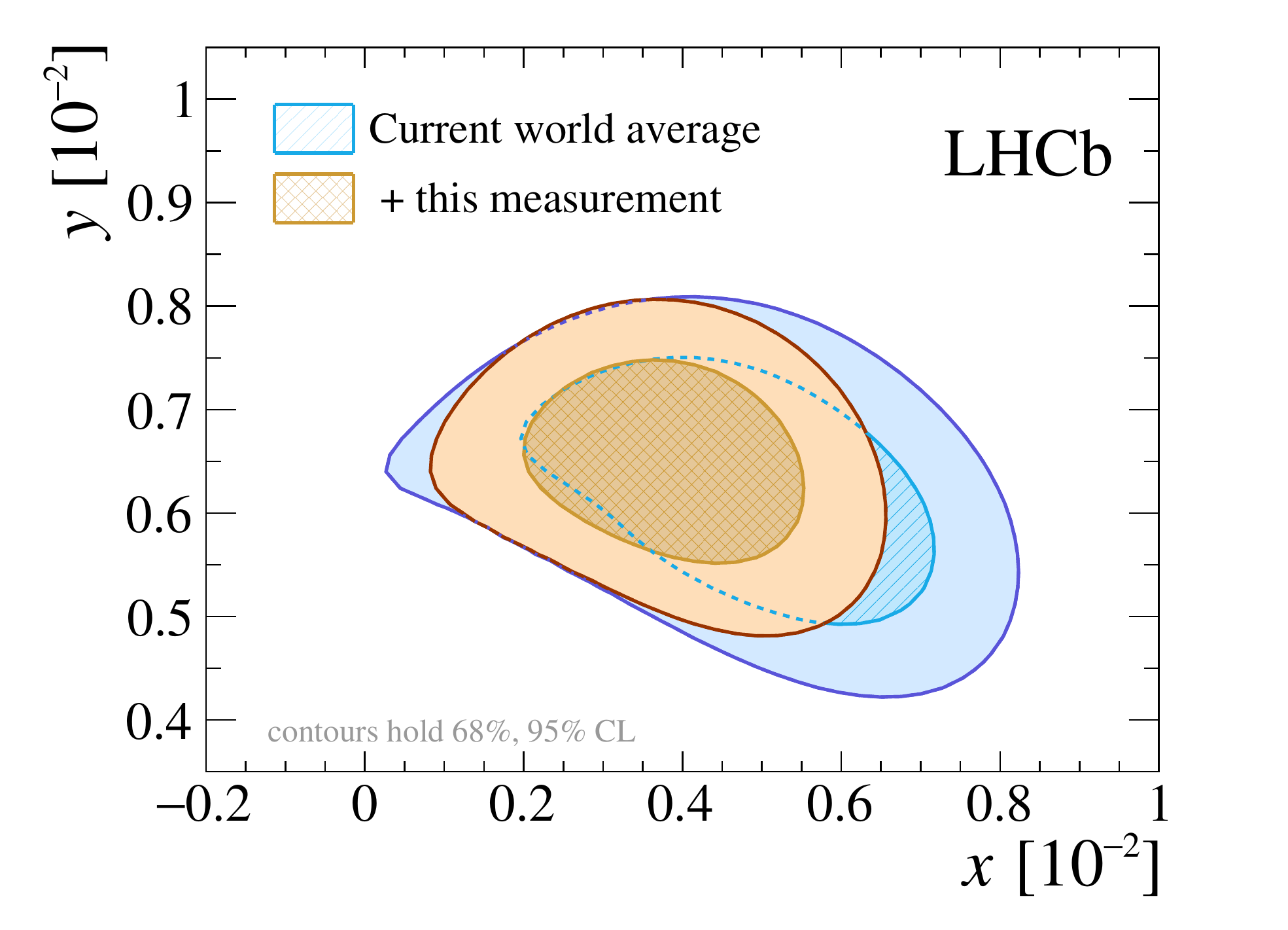}
\includegraphics[width=0.36\textwidth]{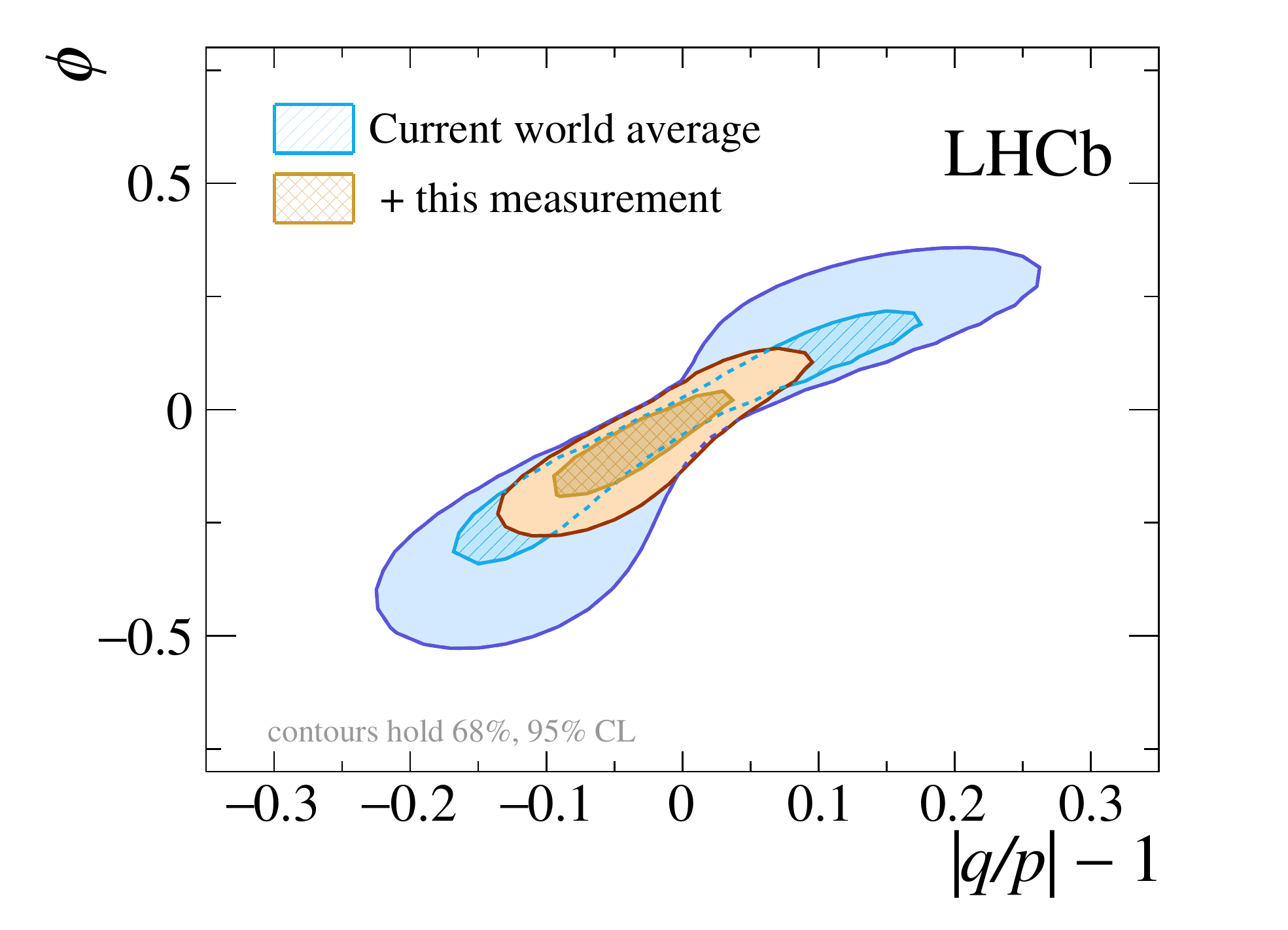}
\caption{Two-dimensional countour plots for (top) $x$ \emph{vs.} $y$ and (bottom) $\phi$ \emph{vs.} $|q/p|-1$ parameters. The blue (orange) shaded areas show the 1 and 2 $\sigma$ level contours for the world average without (with) the inclusion of this measurement.} \label{fig:WAmixing}
\end{figure}

The results obtained are
\begin{eqnarray}
x_{\CP} &=& (\phantom{-}2.7 \pm 1.6 \pm 0.4)\times 10^{-3}, \nonumber \\
y_{\CP} &=& (\phantom{-}7.4 \pm 3.6 \pm 1.1)\times 10^{-3}, \nonumber \\
\Delta x &=& (-0.53 \pm 0.70 \pm 0.22)\times 10^{-3}, \nonumber \\
\Delta y &=& (\phantom{-}0.6 \pm 1.6 \pm 0.3)\times 10^{-3}, \nonumber
\end{eqnarray}
where the first uncertainties are statistical and the second systematic. The results are consistent with the hypothesis of no \CP violation. The resulting determination of the mass difference, as well as those of the \CP-violating parameters, are the most precise from a single experiment. When combining the result for the mass difference with the world average~\cite{Amhis:2016xyh}, the first evidence for a value $x>0$ is achieved. The impact of the parameters measured in this analysis on the current world average is shown in Fig.~\ref{fig:WAmixing}.

\vspace{-4.5cm}
\section{Conclusions}
\vspace{-0.8cm}
In these proceedings, the state of the art for what regards the search for \CP violation in charm meson decays has been presented.
In particular, the results shown include the first observation of \CP violation in charm decays by the LHCb collaboration. In the next decade, the LHCb and Belle 2 collaborations will continue the search for direct and indirect \CP violation in several other decay modes, in order to help clarify the picture and establish whether new dynamics is at play in the up-quark sector.

\bigskip 

\end{document}